# Ransomware IR Model: Proactive Threat Intelligence-Based Incident Response Strategy

Anthony, Lai

VX Research Limited, darkfloyd@vxrl.hk

Hong Kong University of Science and Technology, lct@connect.ust.hk

PING FAN, KE

Singapore Management University, pfke@smu.edu.sg

Alan, Ho

VX Research Limited, alanh0@vxrl.hk

Ransomware impact different organizations for years, it causes huge monetary, reputation loss and operation impact. Other than typical data encryption by ransomware, attackers can request ransom from the victim organizations via data extortion, otherwise, attackers will publish stolen data publicly in their ransomware dashboard forum and data-sharing platforms. However, there is no clear and proven published incident response strategy to satisfy different business priorities and objectives under ransomware attack in detail. In this paper, we quote one of our representative front-line ransomware incident response experiences for Company X. Organization and incident responder can reference our established model strategy and implement proactive threat intelligence-based incident response architecture if one is under ransomware attack, which helps to respond the incident more effectively and speedy.

## 1 INTRODUCTION

We have dealt with ransomware investigations, incident response and analysis for last 5 years. Most of the companies are attacked because of the system is unpatched with remote code execution vulnerabilities, vulnerable remote desktop connection exposed to Internet, staff has downloaded and executed suspicious malware due to phishing email and visit popular software download forum, etc. The attack is that the documents and media files are encrypted in the victim's machine and his/her mounted external storage. Some of our customers have no data backup at all and need to pay the ransom straight away with our assistance to decrypt and restore the data; Some have data backup, unfortunately, their data is extorted and attackers request ransom, otherwise, they will publish stolen data in public data-sharing forum or platform. In this paper, we illustrate the latter type of ransomware attack scenario. We set up an incident response model for organization to reference and implement when they are under ransomware attack. In fact, there are published ransomware incident response playbook[1][2] with technical plan, response task and communication list. Some incident response plan follows standard incident response procedure including Detect-Contain-Remediation [3] and National Institute of Standards and Technology (NIST) publishes a collection of different documents on how to get started, detect, and respond to ransomware threat [4].
However, we have found their limitations:

Firstly, they mainly cover technical attack scenario and forensic investigation instead of business objective, recovery priority and negotiation with ransomware operator. For example, is the priority of data recovery and extorted data handling higher? How can we communicate and negotiate with the ransomware operator and verify what kind of data is extorted?

Secondly, published incident response strategy is not well-proven, commented and executed through real-life intensive ransomware incident response experience like us. We have executed both negotiation and technical incident response strategy successfully. We have attempted to get the best elements of their incident response strategy into our model.

Thirdly, there is recent research [5] about waking up digital forensic community to detect and defend ransomware, from its search results in Google Scholar, it mentions that the case study and incident response plan should be added and contributed more by community to improve the overall ransomware incident response and defense. It is found that there are relatively low number of contributions in incident response strategy and management, it motivates us to contribute our experience and model to the community.

Finally, in our paper, we have set up a technical threat intelligence-based architecture for ransomware incident response in data extortion scenario, which is not published in other publications.

## 1.1 Company X Background

There is a company X which suffered from ransomware attack, and we are handling the investigation and incident response. They are a typical small-to-medium sized enterprise, with around 200-300 end user workstations and around 30-40 servers on premise and in cloud. Those servers are set up for their customers' project individually. Basically, they have an IT team to handle server administration and technical support, however, without security monitoring and ongoing vulnerability update over their servers but with typical anti-virus software installed. In Company X, they have regular online data backup and no offsite data backup, however, customer officer and supervisor will mount those data backup drive even in non-office hour. Fortunately, Company X has subscribed a cyber security insurance in case of any data leakage and attack.

## 1.2 Incident Scenario

We are called as they have found their data and customer documents are encrypted. We have investigated into the incident and found out that one of their staff VPN's account credentials is compromised because their Fortigate VPN gateway is vulnerable [X] without timely security update, it allows attackers access Company X's internal network and execute lateral movement[X] to understand the entire network and figure out the mounted storage with customer and business documents. The ransomware gang, REvil [X], threatens Company X as they claim they extorted more than 1TB of customer data. Company X can still recover most of their data, however, their worries fall on the publishing of extorted customer documents and data, they always question us:

    1. How much data did the attacker extort from our network?
    2. When will the attacker publish the data?
    3. Other than ransomware dashboard provided by attacker, is there any new locations exposing our leaked data?
    4. Can we download all published data and get to know the leak exposure level?
    5. Can we take down the extorted data sharing page or link immediately after we have downloaded the data?

Here are the questions from customers for Company X:
    1. Their customer needs external security company to assess Company X's server is secure or not.
    2. How much data is leaked with our data?
    3. Are our server affected in this incident?
    4. Do you have any incident response plan to handle the same incident in the future?
    5. Do you have or plan to implement technical solutions in Data Loss Prevention, Security Monitoring and End-Point
Threat Detection and Response (ETDR) after incident?

Company X has not paid the ransom, however, after a period of negotiation time, REvil decided to publish the extorted data, we have downloaded the data and proceeded on data link sharing take-down operation. Originally, REvil has decided to put another extorted file, however, it is taken down by Russian government [X].

When everyone thinks the story should be ended, Black Matter [X] announces to be the REvil's successor and request ransom from Company X, otherwise, they publish a larger volume of extorted data. Company X still refuses to pay the ransom. Fortunately, we have found no data is published when the publishing deadline hits. The Black Matter dashboard is taken down by government in a few days after the deadline.

## 2 RANSOMWARE INCIDENT RESPONSE MODEL

We have proposed this model to answer Company X's and their external parties' concern and questions. We are required to design an incident response strategy to address their priorities, objectives, and concern, to customize a suitable incident response plan instead of just dependent on typical incident response methodology. It comprises three sub-models including:
- **Ransomware IR Strategy** – It is a high-level strategy model for decision maker to justify what kind of actions should be done in different ransomware attack scenarios and with various business priorities.
- **Threat Intelligence-Driven IR Technical Architecture** – A technical architecture with corresponding tools and approach for ransomware incident response in data extortion scenario.
- **Strategy Implementation Matrix** – A supplementary matrix on how to implement the strategy.

### 2.1 Ransomware Incident Response (IR) Strategy Model

Here is the strategy table for us to decide which IR strategy we set up for Company X.

Table 1: Ransomware IR Strategy

| | | Data Encryption + Extortion | Data Encryption only | Data Extortion only |
|---|---|---|---|---|
| Data Backup | High value | **Negotiation Strategy:**<br>- We need to know what kind of data extorted by attacker.<br>- Whether the data is high value from victim company and victim company's external parties' perspectives.<br>- Decide whether we pay ransom to stop the high value data publishing or not. If not, we prepare "Extorted Data Publishing Response".<br>**Technical Strategy:**<br>- Identify ransomware attacker group.<br>- Identify which folder(s)/data are encrypted.<br>- Figure out the attack vector of and vulnerabilities taken by ransomware group. If identified, instruct to disable or update security vulnerabilities of the affected systems/services.<br>- Contain the incident and confirm there is data is still extorted and leaked out.<br>- Examine logs in server/network devices and check what kind of data and folder are accessed. | **Negotiation Strategy:**<br>- We do not need to negotiate with ransomware operator.<br>**Technical Strategy:**<br>- Same as technical strategy in "Data Encryption + Extortion with Data Backup" for high value data.<br>- Recover affected data from backup. | **Negotiation and Extorted Data Publishing Response Strategies:**<br>- Same as strategies in "Data Encryption + Extortion with Data Backup" for high value data.<br>**Technical Strategy:**<br>- Same as technical strategy in "Data Encryption + Extortion with Data Backup" for high value data. |

|  |  | **Extorted Data Publishing Response Strategy (\*):**<br>- Check any ransomware dashboard exposing victim's data.<br>- Prepare to download the published data.<br>- Prepare to notify the data sharing platform to take down the data sharing link.<br>\* It is covered by our proposed threat intelligence-based ransomware IR architecture. |  |  |
| --- | --- | --- | --- | --- |
|  | Low value | **Negotiation Strategy:**<br>- We do not need to proceed on negotiation of ransom.<br>- We can attempt to find out what kind of vulnerabilities they have taken to compromise our network.<br>**Technical Strategy:**<br>- Same as technical strategy in "Data Encryption + Extortion with Data Backup" for high value data. | **Technical Strategy:**<br>- Same as technical strategy in "Data Encryption + Extortion with Data Backup" for low value data.<br>- Recover affected data from backup. | **Technical Strategy:**<br>- Same as technical strategy in "Data Encryption + Extortion with Data Backup" for low value data. |
| No Data Backup | High value | **Negotiation Strategy:**<br>- Same as negotiation strategy in "Data Encryption + Extortion with Data Backup" for high value data.<br>**Technical Strategy:**<br>- Include technical strategy in "Data Extorted with Data Backup"<br>- Figure out any published decryptor of the ransomware group.<br>- Attempt to decrypt files with published decryptor.<br>- Consider recovering any data from alternative channels including email and USB storage of staff, if any.<br>**Ransom Payment Strategy:**<br>- If consider paying ransom. Ensure the ransomware attacker can recover your encrypted file via testing. We will cover details in the implementation about the testing.<br>**Extorted Data Publishing Response Strategy (\*):** | **Technical Strategy:**<br>- Figure out any published decryptor of the ransomware gang.<br>- Attempt to decrypt files with published decryptor.<br>- Consider recovering any data from alternative channels including email and USB storage of staff, if any.<br>- Consider paying ransom.<br>- Ensure the ransomware attacker can recover your encrypted file via testing. | **Negotiation and Extorted Data Publishing Response Strategies:**<br>- Same as strategies in "Data Encryption + Extortion with No Data Backup" for high value data.<br>**Technical Strategy:**<br>- Same as technical strategy in "Data Encryption + Extortion with No Data Backup" for high value data. |

|  |  | Data Encryption + Extortion with Data Backup | Data Encryption + Extortion without Data Backup | Data Encryption only without Data Backup | Data Extortion only without Data Backup |
|---|---|---|---|---|---|
|  |  | - Same as extorted data publishing strategy in "Data Encryption + Extortion with Data Backup". |  |  |  |
|  | Low value | **Negotiation Strategy:**<br>- We do not need to proceed on negotiation.<br>**Technical Strategy:**<br>- Same as extorted data publishing strategy in "Data Encryption + Extortion with Data Backup". | **Negotiation Strategy:**<br>- We do not need to proceed on negotiation.<br>**Technical Strategy:**<br>- Same as technical strategy in "Data Encryption + Extortion with Data Backup" for low value data.<br>- Recover affected data from backup. | **Negotiation Strategy:**<br>- We do not need to proceed on negotiation.<br>**Technical Strategy:**<br>- Same as technical strategy in "Data Encryption + Extortion with Data Backup" for low value data. |  |

**Table Structure and Definitions:**
Company with Data backup and no Data Backup, for each row, we have:
- High value data definition: Confidential and/or sensitive data for Company X and their external parties (i.e. customers and contractors).
- Low value data definition: It does not has significant impact over the business operation, reputation, privacy issue and legal obligation.

Three different attack scenarios in columns:
- Data encryption and extortion – The data is encrypted and extorted for ransom request.
- Encryption only – Data is encrypted for ransom request.
- Extortion only – Data is stolen and extorted for ransom request.

For each cell, we will cover strategies:
- Negotiation – Attempt to negotiate with ransomware operator for the best interest of the Company X.
- Technical – Carry out technical incident response, forensics and investigation tasks to confirm the threat impact level, data leak level, attack vector, vulnerabilities manipulated to compromise Company X and contain the incident.
- Ransom Payment – Confirm the ransomware group is capable to recover the encrypted file and/or the decryptor is working if ransom payment is successful.
- Extorted Data Publishing – Another incident response to download the published extorted data and take down the publishing link.

**Best Case Vs Worst Case:**
- Best cases: They always fall over low value data with or without data backup, regardless of data encryption and/or extortion and high value data with data backup without data extortion
- Worst cases: They are about no data backup for high value data (data encryption or/and extortion) and high value data with data backup but data is extorted or/and encrypted.

**Strategy Checklist:**
- We have covered and highlighted the mandatory tasks done under each strategy.
- For the implementation details of the strategy checklist, we have included details in Strategy Implementation Matrix.

## 2.2 Threat Intelligence-Driven Incident Response Architecture

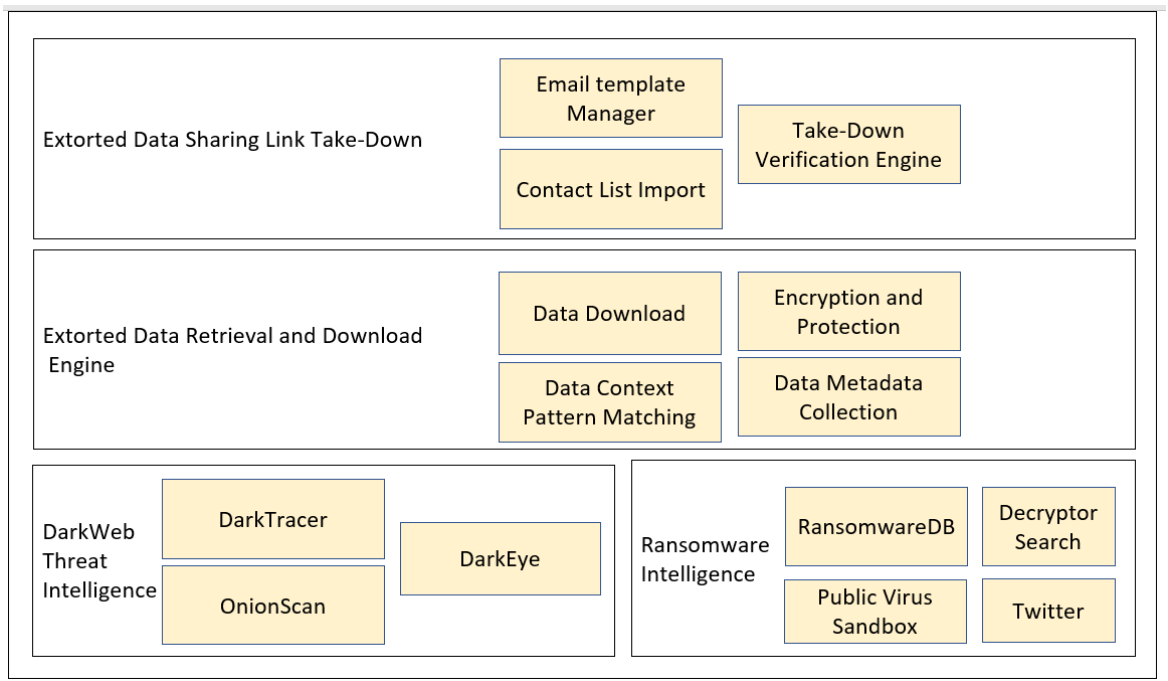

Figure 1. Threat Intelligence-Driven Ransomware Incident Response Architecture

We establish a technical architecture to deal with ransomware incident response in the data extortion scenario as Company X rejects to pay the ransom, it contributes to our monitoring and incident response on continuous basis:

**Dark Web Threat Intelligence**
- This module aims to crawl the darkweb (DarkTracer [X] (https://darktracer.com/) and DarkEye [X] (https://www.knownsec.asia/darkeye/). Check any link and dashboard with Company X's ransomware news and post about the leaked data.
- Alerts will be sent if any links are found. They are sent through email and Telegram.
- Scan the target ransomware dashboard and link for the availability (OnionScan [X])( https://onionscan.org/) and collect technical footprints of the onion address for investigation purpose.

**Ransomware Intelligence**
- Crawl all the current and live ransomware group dashboard for any Company X's data (https://github.com/fastfire/deepdarkCTI/blob/main/ransomware_gang.md)

- Check with latest victims from RansomDB (https://www.ransom-db.com/)
- Search for any latest published decryptor of relevant ransomware group from the following sources: https://www.nomoreransom.org/en/decryption-tools.html / https://github.com / https://twitter.com.
- Check any link and dashboard with Company X's ransomware news and post about the leaked data. For this case, we check over the REvil dashboard at here: http://dnpscnbaix6nkwvystl3yxglz7nteicqrou3t75tpcc5532cztc46qyd.onion
- Identify our captured malware and scan with virus sandbox to confirm those artifacts and behavior (any.run)
- Crawl Twitter (https://twitter.com) for any latest update of ransomware attack and dashboard about Company X.Twitter is favorable by ransomware groups to announce victim.

**Extorted Data Retrieval and Download Engine**
- Download the data from cloud-based server.
- Schedule the download in different period and see whether there is a difference of the download file. Compare and calculate the metadata of the downloaded files. Compare the downloaded data size and hash.
- Apply regular expression and search for sensitive information.
- Extract folder and file name list from downloaded data for verification.
- Encrypt and archive the downloaded extorted data.

**Extorted Data Sharing Link Take-Download**
- Set up email template with take down request
- Import the email and communication list of the data sharing platforms
- Send email and instant messenger message to relevant parties to take down the data sharing link
- Once take down is done, verify the download link is still available or not
- Some solicitor firm with cyber security and ransomware handling experience can undertake the above task.

### 2.3 Strategy Implementation Matrix

From the below table, we have listed out "how-to" details which cannot be covered in our above strategy table.

Table 2: Ransomware IR Break Down

| Strategy | Implementation Description of Company X |
|---|---|
| **Negotiation Strategy:**<br>- We need to know what kind of data extorted by attacker.<br>- Whether the data is high value from victim company and victim company's external parties' perspectives.<br>- Decide whether we pay ransom to stop the high value data publishing or not. If not, we prepare "Extorted Data Publishing Response". | - We discussed with Company X and checked which files are important to recover.<br>- Negotiate with operator, request operator to decrypt some of our sample files before sending ransom. Verify the decryption is successful<br>- Check with Company X whether they are willing to pay the ransom or not. In this case, they refuse to pay the ransom.<br>- Throughout the negotiation, we attempt to delay the publishing data schedule. i.e. getting approval from management<br>- Negotiating a lower ransom price i.e. because of business context. |
| | - Negotiate with operator to disclose the vulnerabilities details if ransom is paid. |
| **Technical Strategy:**<br>- Identify ransomware attacker group.<br>- Identify which folder(s)/data are encrypted.<br>- Figure out the attack vector of and vulnerabilities taken by ransomware group. If identified, instruct to disable or update security vulnerabilities of the affected systems/services.<br>- Contain the incident and confirm there is data is still extorted and leaked out.<br>- Examine logs in server/network devices and check what kind of data and folder are accessed.<br>- Figure out any published decryptor of the ransomware group.<br>- Attempt to decrypt files with published decryptor.<br>- Consider recovering any data from alternative channels including email and USB storage of staff, if any. | - In this case, we have checked every logs of network/security/server devices, we can identify the unauthorized access but cannot confirm how much data is extorted.<br>- Figured out VPN is vulnerable; Instruct Company X to stop the corresponding service.<br>- Mandate to change the compromised account's password<br>- Mandate to reset password for all staff; Implement 2FA for critical servers including domain controller and servers in cloud.<br>- Mandate to install latest Windows, VPN and backup storage devices security patches.<br>- Scan all affected machines with additional two anti-virus detector, there is no further ransomware found in the network.<br>- Decrypt those affected encrypted files with published decryptors, however, we failed. |
| **Ransom Payment Strategy:**<br>- If consider paying ransom. Ensure the ransomware attacker can recover your encrypted file via testing. We will cover details in the implementation about the testing. | In this case, Company X refused to pay the ransom:<br>- Delay the publishing schedule i.e. claim to getting approval.<br>- Negotiate lower ransom.<br>- The REvil attacker group has published the data once, however, we can take them down with our architecture. Afterwards, REvil and their successor Black Matter have been taken down by Russian governmental action (https://threatpost.com/russian-security-revil-ransomware/177660/) [X]. No more extortion ransom is requested.<br><br>If prepare to pay ransom:<br>- Prepare a bitcoin wallet with bitcoin.<br>- Attempt to pay the ransom partially, and send the partial data for decryption, those partial data is critical to business.<br>- Claim to pay the remaining ransom if the data can be decrypted, this is business management suggestion.<br>- Can repeat to pay partial ransom and request decryption of critical files.<br>- If the business owner agrees to pay the entire ransom, we are responsible for testing the decryptor is working.<br>- Keep close communication with the ransomware operator in chatroom, normally, they have a chatroom to communicate with the victim, ensure all technical requirement and operation of decryptor are satisfied and working<br>- |
| **Extorted Data Publishing Response Strategy (*):**<br>- Check any ransomware dashboard exposing victim's data. | - The implementation is mainly supported by our threat intelligence-based ransomware IR architecture. |
| - Prepare to download the published data.<br>- Prepare to notify the data sharing platform to take down the data sharing link. | For Company X case:<br>- Set up cloud-based server in AWS to download the published data.<br>- Prepared another separated debit card to purchase the download accelerator<br>- Purchase the download accelerator to make sure we download the data faster.<br>- We have set up channels to communicate the data sharing platform to prepare to take down the data publishing link.<br>- We need to check the taken-down sharing link is regularly. Somehow, it will be live again, we are required to alert the business owner.<br>- We need to monitor the dark web for any data extortion posted by other ransomware group. It is possible other group takes over the victim data from another. |

## 3 RELATED WORK

This section cites a variety of journal [5, 15], conference [1, 6, 8, 12, 13], and magazine [3] articles to illustrate how they appear in the references section. It also cites books [9, 10], a technical report [7], a PhD dissertation [4], an online reference [14], a software artifact [11], and a dataset [2].

As you build your article, you should note where you will be placing citations. If you are using numbered citations and references, the reference number - "...as shown in [5]..." is sufficient. If you are using the "author year" style, a reasonable placeholder is the primary author's last name and the year of publication - "...as shown in [Harel 1978]..." - we will be updating this placeholder later in the process with the citation label as generated by the Word macros in the "master template.

## 4 CONCLUSION

In this paper, we have walkthrough our ransomware incident response strategy illustrated with a case study of Company X. The proposed strategy and architecture are decided on business priority and incident response objective, it aims to save resources of organizations and incident responder on ransomware incident response, keeping the business continuity and recovering the operation from attack. Furthermore, we cannot assume the ransomware group will not just "hit and go", it is possible for them to share the victim vulnerabilities and extorted data with another ransomware group, organizations must not relax even after their data is restored or data value is low, they should figure out all possible attack vectors and fixed vulnerabilities to prevent from revisit of ransomware attack.